\documentclass[12pt,letterpaper]{article}

\usepackage{amscd,amsmath,amssymb,amsfonts,xspace,mathrsfs}
\usepackage{slashed}
\usepackage{graphicx}
\usepackage{tikz-cd}
\tikzset{>=latex} 
\tikzcdset{arrow style=tikz}
\usetikzlibrary{shapes.geometric,positioning}
\usetikzlibrary{bending} 
\usetikzlibrary{shapes}  
\usetikzlibrary{calc}   
\tikzset{stretch/.initial=1}
\newcommand\drawloop[4][]%
   {\draw[shorten <=0pt, shorten >=0pt,#1]
      ($(#2)!\pgfkeysvalueof{/tikz/stretch}!(#2.#3)$)
      let \p1=($(#2.center)!\pgfkeysvalueof{/tikz/stretch}!(#2.north)-(#2)$),
          \n1= {veclen(\x1,\y1)*sin(0.5*(#4-#3))/sin(0.5*(180-#4+#3))}
      in arc [start angle={#3-90}, end angle={#4+90}, radius=\n1]%
   }

\usepackage{hyperref}

\usepackage[multiple]{footmisc}
\usepackage{url}

\numberwithin{equation}{section}

\def\varpi{t}

\def\det{\,{\rm det}\, }

\newcommand{\p}{\partial}

\def\({\left(}
\def\){\right)}
\def\[{\left[}   
\def\]{\right]}
\def\<{\left\langle}
\def\>{\right\rangle}

\newcommand{\CF}{\mathcal{F}}

\newcommand{\CG}{\mathcal{G}}

\newcommand{\CM}{\mathcal{M}}

\newcommand{\CN}{\mathcal{N}}
\newcommand{\CE}{\mathcal{E}}
\newcommand{\CX}{\mathcal{X}}
\newcommand{\CP}{\mathcal{P}}

\newcommand{\CT}{\mathcal{T}}

\DeclareSymbolFont{AMSa}{U}{msa}{m}{n}
\DeclareSymbolFont{AMSb}{U}{msb}{m}{n}
\DeclareMathSymbol{\fieldR}{\mathalpha}{AMSb}{"52}

\newcommand{\CO}{\mathcal{O}}

\newcommand{\Tr}{\mbox{Tr}}

\def\bea{\begin{eqnarray}}
\def\eea{\end{eqnarray}}
\def\be{\begin{equation}}
\def\ee{\end{equation}}
\def\ba{\begin{align}}
\def\ea{\end{align}}
\def\bse{\begin{subequations}}
\def\ese{\end{subequations}}

\newcommand{\bfk}{{\boldsymbol k}}

\newcommand{\bfp}{{\boldsymbol p}}

\newcommand{\bfx}{{\boldsymbol x}}

\def\ba{\bar a}

\newcommand{\CL}{{\cal{L}}}
\newcommand{\CA}{{\cal{A}}}

\def\cij#1{c}
\def\ci#1{c}

\def\gamD#1{\tilde\gamma}

\def\cl0{\tilde c_0}

\newcommand{\bfmu}{{\boldsymbol \mu}}
\newcommand{\bfnu}{{\boldsymbol \nu}}

\newcommand{\noi}{\noindent}

\begin{document}

\begin{titlepage}

\begin{center}

\hfill  
{}
\vskip 2cm {\LARGE \bf Four-Manifold Invariants and Donaldson-Witten Theory \\ \vskip .4cm } \vskip
1.25 cm { Jan Manschot}\\
\vskip 0.5cm  
{\it School of Mathematics, Trinity College, Dublin 2, Ireland}\\
\vskip .3cm
{\it Hamilton Mathematical Institute, Trinity College, Dublin 2,
  Ireland}\\
\vskip .3cm
{\it School of Natural Sciences, Institute for Advanced Study, 1
  Einstein Drive, Princeton, NJ 08540 USA}

\end{center}

\vskip 2 cm

\begin{abstract}
\baselineskip=18pt 
\noi  This article surveys invariants of
four-manifolds and their relation to Donaldson-Witten theory, and other topologically twisted Yang-Mills
theories. The article is written for the second edition of the Encyclopedia of
Mathematical Physics, and focuses on the period since the first
edition in 2006. \\

\end{abstract}

\end{titlepage}  
 
\pagestyle{plain} 
\baselineskip=19pt

\section{Introduction}
A hallmark of mathematical physics is the relation between geometry
and gauge theory. On the one hand, the geometric formulation of gauge
theory gives profound insights, while gauge-theoretic observables have proven
crucial for the understanding of the underlying space-time geometry. 
Donaldson-Witten theory is one of the purest examples of this
connection in four dimensions. This theory is a topologically
twisted gauge theory, and well suited to 
exact results and connections to mathematics. The theory was formulated by
Witten in 1988 \cite{Witten:1988ze}, following developments
by Donaldson in four-manifold geometry
\cite{DONALDSON1990257}. These were in turn inspired by 
non-perturbative solutions \cite{Belavin:1975fg} in
Yang-Mills theory \cite{Yang1954}. The subject
thus draws heavily on both physics and mathematics, which has been reviewed in
the 2006 version of the Encyclopedia of Mathematical Physics
\cite{BAUER2006457, MARINO2006110, NASH2006386}.

This review will mostly focus on developments since
2006.  To this end, Donaldson-Witten theory is interpreted broadly to include
any four-dimensional topological quantum field theory obtained as a topological twist of a
quantum field theory with extended supersymmetry. Different
topologically twisted theories give rise to different invariants of
gauge theoretic moduli spaces \cite{Vafa:1994tf, Losev:1995cr, Dijkgraaf:1996tz, Moore:1997pc,
  LoNeSha}. Recent years have seen much progress on the rigorous
formulation of these invariants as well as explicit determination using
mathematical and physical approaches.

Section \ref{FManClass} reviews the classification of four-manifolds
and various invariants of moduli spaces related to four-manifolds.
Sections \ref{TTYM} and \ref{N2theories} discuss twists of
  supersymmetric gauge theories and their relation to different mathematical invariants
  reviewed in Section \ref{FManClass}. Section \ref{Secuplane} reviews
  recent developments on the formulation and
  evaluation of the $u$-plane integral, which is a physical approach
  to Donaldson invariants \cite{Witten:1995gf, Moore:1997pc}.

\section{Four-manifolds and their classification}
\label{FManClass}
 
\subsection{Basic aspects of four-manifolds}
\label{4manBasic}
Basic topological invariants of an oriented, closed four-manifold $X$ are its Euler number $\chi(X)$ and signature
$\sigma(X)=b_2^+(X)-b_2^-(X)$. We let the lattice $L$ be the
image of $H^2(X,\mathbb{Z})$ in $H^2(X,\mathbb{R})$. The intersection
form on $H^2(X,\mathbb{Z})$ provides a natural non-degenerate,
bilinear, uni-modular
form,
\be
\label{BX}
B_X: L\times L \to \mathbb{Z},
\ee 
which can be extended by linearity to $L\otimes \mathbb{R}$ and
$L\otimes \mathbb{C}$.
For $b_2^+=1$, the period point $J\in L\otimes \mathbb{R}$ is the unique point in the forward
lightcone of $L$, such that $J=*J$ and $J^2=1$, with $*$ the Hodge star.
 
If $X$ admits a Spin structure, $L$ is an even lattice. Every oriented
four-manifold admits a Spin$^c$ structure. For later reference, recall
that Spin$^c(4)$ is defined as the group
\be
{\rm Spin}^c(4)=\left\{(u_1,u_2)\vert \det(u_1)=\det(u_2) \right
\}\subset U(2)\times U(2).
\ee
A ${\rm Spin}^c$ structure $\frak{s}$ is a principal Spin$^c(4)$
bundle compatible with the principal $SO(4)$ bundle associated to the
tangent bundle $TX$. 
The fundamental two-dimensional representation of $U(2)$ gives two
inequivalent representations of Spin$^c(4)$ corresponding to the two
projections of Spin$^c(4)$ to $U(2)$. The associated bundles $W^\pm\to X$ are the chiral spin bundles. We define the line bundle $\CL=\det(W^{\pm})$. The first Chern class of
$\CL$ is the characteristic class $c(\frak{s})$ of the Spin$^c$
structure $\frak{s}$, which takes values in $\bar w_2(X) \mod 2L$,
with $\bar w_2(X)$ a lift of the Stiefel-Whitney class $w_2(X)$ to $L$. 

If $X$ is an almost complex four-manifold, the real tangent bundle $TX$ has
a complex structure, such that one can choose a metric for which the
 structure group of $TX$ is reduced to $U(2)$. This gives rise to a canonical Spin$^c$
structure for such four-manifolds, since $U(2)$ embeds naturally in Spin$^c(4)$ \cite{Manschot:2021qqe}. Thus transition
functions for $TX$ are valued in $U(2)$, and define
Spin$^c(4)$ valued transition functions. This is the canonical
Spin$^c(4)$ structure associated to an almost complex structure.

\subsection{Classification}
The classification of four-manifolds has been a leading objective in
mathematics with an impact on a wide range of subjects. There are various ways for
this classification. First, we can consider classifying four-manifolds
up to {\bf homotopy}. A classic result for this classification, is
that a simply connected, closed and oriented four-manifold $X$ is classified up
to homotopy by its intersection form $B_X$ (\ref{BX}) \cite{Whitehead1949, Milnor1958}. 
  
Next, we can consider classification up to
{\bf homeomorphism}. Freedman \cite{Freedman1982} proved that for simply connected
compact four-manifolds this classification is determined by the
intersection form and the Kirby-Siebenmann invariant, ${\rm ks}(X)\in H^4(X,\mathbb{Z}_2)$. For a given {\it even} uni-modular
intersection form $B_X$, ${\rm ks}(X)$ is determined by
$B_X$,
\be 
\label{ks} 
{\rm ks}(X)=\frac{1}{8} \sigma(X) \mod 2. 
\ee
There exists therefore only one simply connected, topological
four-manifold with that intersection form. If the intersection form is {\it odd}, there
exists two inequivalent four-manifolds distinguished by their value of
${\rm ks}$. For indefinite lattices, Serre's classification
theorem asserts that there is a unique lattice up to isomorphism given the signature, dimension
and parity. On the other hand, for the number of positive or negative
definite, unimodular lattices, only a lower bound is
known. 

Much more intricate and more relevant for the connection to Donaldson-Witten theory is the classification of four-manifolds with a
smooth $C^\infty$ structure. Two smooth structures on a topological manifold $X$
are said to be equivalent if related by a {\bf diffeomorphism}. It is
argued that there is no algorithm which can determine whether two
arbitrary closed four-manifolds are diffeomorphic,
because the isomorphism problem for finite group presentations is
undecidable \cite{Markov:1958}. This issue is avoided if we fix the fundamental group. In this paper, we will typically consider
simply connected four-manifolds, $\pi_1(X)=1$.

Not all bilinear forms occur as the intersection form of a compact,
oriented, {\it smooth} four-manifold.
There are then two main questions:
\begin{enumerate}
\item The {\bf geography problem}: Which bilinear forms $B$ are realised as
  the intersection form $B_X$ of a simply connected, compact, smooth
  four-manifold $X$?
\item The {\bf botany problem}: If $X$ admits a smooth structure, how many
  inequivalent smooth structures does $X$ admit?
\end{enumerate}

A necessary, but not sufficient, condition for $X$ to have a smooth
structure is that ${\rm ks}(X)$ vanishes. Therefore, at least one
of the two four-manifolds mentioned below (\ref{ks}) with $B$ odd, is not smooth.  
Moreover, (\ref{ks}) demonstrates that $\sigma(X)=0\mod 16$ for a
simply connected, smooth four-manifold, while the signature of a generic even lattice 
vanishes modulo 8. This is also known as Rokhlin's theorem.

Another remarkable result for the geography problem is Donaldson's theorem, which
asserts that if $B_X$ is positive or negative
definite, it is necessarily diagonalizable \cite{Donaldson1983}. Part of the importance of
this result is that it's proof is based on instanton moduli spaces, and thus
a first instance where Yang-Mills theory is relevant for
four-dimensional topology. 
   
We further mention the 11/8 conjecture which states that for compact four-manifolds with an even
intersection form, a smooth structure requires \cite{Matsumoto:1982}
\be 
b_2(X)\geq \frac{11}{8}|\sigma(X)|.
\ee 
Furuta \cite{Furuta2001} has given a proof for the weaker inequality $b_2(X)\geq \frac{10}{8}|\sigma(X)|$.
Recent progress is made on intermediate cases \cite{Hopkins2018}.
  
The understanding for the botany problem remains far less
complete. Donaldson polynomial invariants relate to this problem since
these were developed to distinguish between
smooth structures \cite{Donaldson90}. These invariants are
introduced in the context of instanton moduli spaces and will play a prominent role
in later sections.

To conclude this subsection, it is worth noting that complex surfaces are partly classified by the
Enriques-Kodaira classification \cite{Barth}. Moreover, manifolds with the
topological condition $b_2^+(X)=1$ play a distinguishing role from the
physical perspective.  The set of such four-manifolds overlaps partly
with the complex surfaces. For further information on such four-manifolds, we
refer to \cite{mcduff:1996} and \cite{park_2004}.

\subsection{Instanton moduli spaces and Donaldson/Seiberg-Witten invariants}
\label{GTIFM}

This subsection gives a brief overview of instanton moduli spaces and
Donaldson invariants. We let $P$ be a principal bundle
for the gauge group $G$. $G$ can be any simple Lie group, but we will
typically set $G=SU(2)$ or $SO(3)$. We furthermore let $\mathcal{\CA}={\rm
  ad}(P)\otimes \mathbb{C}$ be the complex vector bundle associated to
the adjoint representation, and $\mathcal{\CE}={\rm
  sp}(P)\otimes \mathbb{C}$ the complex vector bundle associated to
the two-dimensional, spinorial representation of the Lie algebra.

For gauge group $SU(2)$ or $SO(3)$, the instanton
number is defined as
\be
k(P)=-\frac{1}{8\pi^2} \int_X {\rm Tr}_{\bf 2}(F\wedge F),
\ee 
where the trace is in the 2-dimensional, fundamental representation of
the Lie algebra. In the following, we will abbreviate ${\rm Tr}_{\bf 2}={\rm
  Tr}$. Not all $SO(3)$ bundles can be lifted to an $SU(2)$
bundle. The obstruction is a non-trivial second Stiefel-Whitney class
$w_2(P)\in H^2(X,\mathbb{Z}/2\mathbb{Z})$,
\be
w_2(P) =\frac{1}{2\pi} {\rm Tr}(F) \mod H^2(X,2\mathbb{Z}).
\ee
An $SU(2)$ bundle with non-trivial $w_2(P)$ is also known as an $SU(2)$ bundle with 't Hooft twisted boundary
conditions \cite{HOOFT1981455}. Moreover, in terms of the terminology
of generalized global symmetries, $w_2(P)$  measures the background
flux for the magnetic 1-form symmetry associated to the center of $SO(3)$ \cite{Gaiotto:2014kfa}. For later use, we also introduce
\be
\label{bfmu}
\bfmu=\frac{1}{2}\bar w_2(P) \in L/2,
\ee
with $\bar w_2(P)$ a lift of $w_2(P)$ to the lattice $L$.

Lifting the bundle $P$ to a complex rank
two vector bundle on an almost complex manifold, the topological
properties of the gauge bundle determine the Chern classes $c_j(\CE)\in H^{2j}(X,\mathbb{Z})$ of $\CE$,
\be
c_1(\CE)= \frac{1}{2\pi} {\rm Tr}(F) ,\qquad k(P)=2\Delta(\CE)=\int_X c_2(\CE)-\frac{1}{4}c_1(\CE)^2.
\ee
 
The instanton moduli space $\CM_k$ is (to first approximation) the space of anti-self-dual connections
modulo gauge transformations $\CG$,
\be
\CM_k=\left\{ A\in \Omega^1(X,\frak{g})\,|\, F=-*F  \right\}/\CG.
\ee
The virtual real dimension of $\CM_k$ is
\be
\label{dimMk}
\dim_{\mathbb{R}}(\CM_k)=8k-3(1-b_1+b_2^+).
\ee
Our interest is mostly in four-manifolds for which
$\dim_{\mathbb{R}}(\CM_k)$ is even, and
thus $b_1+b_2^+$ odd. A change of the metric of $X$
changes $\CM_k$ by a bordism. Thus cobordism invariants, such as the
Donaldson invariants, are topological invariants of $X$
\cite{DONALDSON1990257}, at least if $b_2^+(X)>1$.

There are at least two aspects to be
addressed for a proper definition of the moduli space and its
topological invariants:
\begin{enumerate}
\item the non-compactness arising from the ``size modulus'' of the
  instanton. That is to say, there exists an infinite sequence of
  instanton solutions in which the curvature of the instanton is
  increasingly concentrated near the center of the instanton.
\item singularities arising from reducible connections, for which the connection one-form $A$
  becomes block diagonal,    
  \be 
A=\left( \begin{array}{cc} A_1 & 0 \\ 0 & A_2 \end{array} \right).
  \ee 
\end{enumerate}  
The first aspect can be dealt with by adding pointlike instantons to
$\CM_k$ as the limiting point of the infinite sequence.

Advanced techniques are required to deal with the second aspect. 
There are roughly two approaches, 1) using geometric analysis
\cite{Uhlenbeck:1982}, or 2) using algebraic geometry for complex algebraic surfaces.
An important result for complex four-manifolds is the correspondence between the
self-duality equation, $F=-*F$ and holomorphic poly-stable vector
bundles and sheaves \cite{Kobayashi1982, Donaldson:1985zz,
  Uhlenbeck1986}. Moduli spaces can then be introduced more formally
in terms of semi-stable coherent sheaves \cite{HuyLen10}.

Having introduced the instanton moduli space $\CM_k$, we can discuss
Donaldson's polynomial invariants. 
The Donaldson map, $\mu_D$, is a map from the homology of $X$ to the
cohomology of the instanton moduli space,
\be
\label{muD}
\mu_D: H_*(X) \to H^{4-*}(\CM_k). 
\ee
If available, the universal bundle $\CP \to \CM_k\times X$ is a useful notion to
understand the Donaldson classes $\mu_D$. The restriction of $\CP$ to
a point in $\CM_k$ equals the bundle $P\to X$. The map $\mu_D$ is then given
in terms of the slant product $/: H^q(X\times \CM) \times H_p(X)\to  H^{q-p}(\CM)$,
\be
\mu_D({\boldsymbol \sigma})=\frac{1}{8\pi^2 } {\rm Tr}[ \CF\wedge
\CF]/{\boldsymbol \sigma},
\ee
with $\CF\in \Omega^2(\CM_k\times X,\frak{g})$ the curvature of $\CP$.
For a two-dimensional submanifold $\Sigma\in H_2(X,\mathbb{Z})$, another interpretation of the image $\mu_D(\Sigma)$ is as the first Chern class of the
determinant line bundle of the virtual Dirac
index bundle over $\Sigma$ \cite[Section 5.2]{Donaldson90}.

The Donaldson invariants are intersection numbers of the images of
$\mu_D$. For $\bfp_i\in H_0(X,\mathbb{Z}), i=1,\dots,\ell$ and $\bfx_j\in
H_2(X,\mathbb{Z}), j=1,\dots,s$, we have
\be
\label{Dint}
D^k(\{\bfp_i\}_\ell, \{\bfx_j\}_s)=\int_{\CM_k} \prod_{i=1}^\ell \mu_D(\bfp_i) \prod_{j=1}^s \mu_D(\bfx_j) 
\ee
Clearly, for this to be non-zero $4\ell+2s=\dim_{\mathbb{R}}\CM_k$,
such that we can form the polynomial
\be
\sum_{\ell,s} D^k(\{\bfp_i\}_\ell, \{\bfx_j\}_s)\, v^\ell\, w^s.
\ee

Witten provided a physical interpretation of the Donaldson
invariants as correlation functions in a topological Yang-Mills
theory \cite{Witten:1988ze}. The physical perspective also gave way to a physical
explanation of the structure theorem by
Kronheimer-Mrowka \cite{Kronheimer1995} for Donaldson invariants, in 
terms of Seiberg-Witten (SW) invariants
\cite{Witten:1994cg}. SW invariants are a count of the solutions to
the Abelian monopole equations,
\be
\label{SWeqs} 
\begin{split}
&  F^+_{\mu\nu}+\frac{i}{2} M \Gamma^{\alpha \dot \beta}_{\mu\nu} \bar
M=0, \\
& \slashed{D} M=0.    
\end{split} 
\ee 
SW invariants in a sense capture the
same information as the Donaldson invariants, but in a more efficient
way. A four-manifold is of SW simple type if the SW invariants vanish if
the virtual dimension of the monopole moduli space vanishes. The
Witten conjecture asserts that for a four-manifold of SW simple type, the
Donaldson invariants are determined in terms of the SW invariants in
an explicit way \cite[Eq. (2.17)]{Witten:1994cg}. The conjectured equivalence is proven for algebraic surfaces in \cite{Gottsche:2010ig}. 
Modulo a few conditions, it is proven for general four-manifolds of SW
simple type in \cite{Feehan_2015}.

SW invariants have been instrumental for many results in geometry and topology.
For example, Taubes demonstrated that if $X$ is
a symplectic manifold, the SW invariants are equivalent to the enumerative
geometry of pseudo holomorphic 2-manifolds \cite{Taubes:1995}.
Moreover, Fintushel and Stern established many smooth structures \cite{Fintushel1998} using SW invariants.
Also, SW invariants vanish for four-manifolds which are
connected sums $X=X_1\# X_2$ with $b_2(X_j)\geq 1$. On the one hand
this is a useful criterion to identify such four-manifolds, while it
also demonstrates that SW invariants are unable to distinguish among
all topological four-manifolds and their smooth structures.

\subsection{Invariants of moduli spaces}
\label{InvModSpace}
Besides their important role for four-manifold geometry, moduli spaces and their topological invariants
have been studied intensely for its own merit. A rather rigorous
framework for moduli spaces has been
developed, which can also be applied to other moduli problems, such as
quiver representations, semi-stable vector bundles on Riemann
surfaces, and Hitchin systems. The rigorous mathematical results also provide a
confirmation of aspects of Yang-Mills theory, such as its path
integral and strongly coupled phenomena.

Especially for complex algebraic surfaces the issues of
non-compactness and singularities can be rigorously dealt with 
using moduli spaces of semi-stable coherent sheaves. Recent years
have seen much mathematical activity related to moduli spaces for algebraic
surfaces and the relation to Donaldson-Thomas theory \cite{Kontsevich:2008fj,
  Joyce:2004tk, Manschot:2016gsx}.  An important
element of the theory is the virtual fundamental class and perfect
obstruction theory \cite{Behrend_1997}. 
Mochizuki's monograph \cite{Mochizuki}
 has developed the intersection theory for moduli 
spaces of instantons on algebraic surfaces, in particular the virtual fundamental class $[\CM_k]^{\rm
  vir}$. Using these techniques it has been possible
to obtain explicit results for various topological invariants of
instanton moduli spaces, notably by G\"ottsche, Kool and
collaborators. See also \cite{Gottsche:2020ale} and \cite{joyce2021enumerative}. It is not
established generally though that these invariants of $\CM_k$ are
also topological invariants of $X$.      
   
We continue with a brief overview of the various invariants of $\CM_k$.  
\subsubsection*{Donaldson invariants}
These were introduced in Eq. (\ref{Dint}). Already in the early 1990's, algebraic surfaces were instrumental to
obtain explicit results for Donaldson invariants, in particular in the
work by G\"ottsche and collaborators \cite{Gottsche1990,
  ellingsrud1995wall, Gottsche:1996aoa, Gottsche:1999}.
For toric manifold $X$, the toric action can be lifted to the moduli space of
  instantons. Consequently, Donaldson invariants can be refined with
  equivariant parameters for the toric symmetry \cite{Gottsche:2006tn,
    Nekrasov:2003vi}. Using localization,
  partition functions are derived for $\mathbb{C}^2$ \cite{Ne, NekOk},
  and also compact toric
  four-manifolds \cite{Nekrasov:2003vi, Gottsche:2006tn, Bershtein:2015xfa}.

\subsubsection*{Euler numbers of instanton moduli spaces}
The Euler number of a
  smooth, closed manifold $M$ is
  commonly defined as,   
\be   
\label{EulerNumber} 
\chi(M)=\int_{M} e(M),
\ee
with $e(M)$ the Euler class of $M$. Singularities and non-compactness
of $\CM_k$ require a more involved definition. Within algebraic geometry, a suitable notion of Euler characteristic
of the moduli space of instantons, or semi-stable coherent sheaves, is
defined, which is a topological invariant of $X$. The definition reads \cite{Gottsche:2018meg, 
    Gottsche:2021dye}   
\be   
\chi^{\rm vir}(\CM_k)=\int_{[\CM_k]^{\rm vir}} c(T^{\rm vir}\CM_k),
\ee
with $c(T^{\rm vir}\CM_k)$ the top Chern class of the virtual holomorphic tangent bundle
of $\CM_k$. Ref. \cite{Fantechi_2010} provides
definitions of the virtual topological Euler characteristic and other
topological invariants in this setting. With the work
\cite{Jiang:2020prk}, these can be extended to arbitrary
choices of first Chern classes of the vector bundle, not only the algebraic classes.
  
A representative example for manifolds with $b_2^+=1$ is the generating function of Euler
characteristics of rank 2 semi-stable, coherent sheaves $\CE$ with $c_1(\CE)=0$ on
$\mathbb{P}^2$ \cite{Klyachko1991, Yoshioka1994, Yoshioka:1995,
  Vafa:1994tf}. This takes the form 
\be
\label{ZP2}
Z_{2}(\tau)=\frac{\sum_{k\geq 0}
  3H(4k)\,q^{k}}{\eta(\tau)^6},\qquad q=e^{2\pi i\tau},
\ee 
where $H(n)$ is the Hurwitz class number and $\eta$ the Dedekind eta
function. In recent years, generating
functions for higher rank sheaves are derived using toric
localization \cite{Kool:2014, weist2009torus} and wall-crossing
\cite{Manschot:2010nc, Mozgovoy:2013zqx, Manschot:2014cca}.
     
For algebraic surfaces with $b_2^+(X)>1$, generating functions of the Euler
characteristics of instanton moduli spaces appear to take the form
of a product of universal
functions of $\tau$, independent of the four-manifold $X$. These universal
functions are multiplied by SW invariants and exponentiated by
elementary invariants such as $\chi(X)$ and $\sigma(X)$. Thus for algebraic
surfaces, these generating functions do not provide further
information for the classification. Explicit expressions for  
the universal functions are conjectured for rank 2 in
\cite{Vafa:1994tf, Dijkgraaf:1997ce}, rank 3 in \cite{Gottsche:2021dye},
and ranks 4, 5 in \cite{Gottsche:2021dye}. The proof that the contribution from the
instanton branch can be expressed generally in terms of universal
functions is expected in upcoming work by Joyce \cite{joyce2021enumerative,
  Joyce:2023, ToAppearJoyce}. 

An important part in this development are generating functions for
Euler charactistics of the monopole branch. This branch is based on
differential equations more similar to \eqref{SWeqs}, and the rigorous
theory was laid out in \cite{Tanaka:2017jom,
  Tanaka:2017bcw}. Using localization with respect to a $\mathbb{C}^*$
action on the moduli space, the Euler numbers of this branch are shown
to be related to nested Hilbert schemes. Laarakker \cite{Laarakker_2020} proved that the  
generating functions for this branch are also composed of universal functions. Coefficients of the
universal functions can be determined term by term through integrals
over Hilbert schemes of points $X^{[n]}$ after localising to a
$\mathbb{C}^*$ fixed locus \cite{Tanaka:2017jom,
  Tanaka:2017bcw, Gholampour:2017bxh}. Using the first
coefficients, one can then make conjectures for the closed expressions
for the universal functions in terms of modular forms. Using
the physical notion of $S$-duality, it is straightforward to conjecture the universal functions for the
instanton branch through the $SL_2(\mathbb{Z})$ action on $\tau$.

\subsubsection*{K-theoretic invariants}
K-theoretic invariants are a class of invariants defined for algebraic surfaces
\cite{Gottsche:2006bm}. Without insertion of Donaldson classes, this
invariant corresponds to the holomorphic Euler characteristics of a
determinant line bundle $\CL \to \CM_k$, 
\be
\label{chihol}
\chi_{\rm hol}(\CM_k, \mu_D(L))=\int_{[\CM_k]^{\rm vir}} {\rm Td}(\CM_k)\,e^{\mu_D(\CL)},
\ee
where $\CL$ is a line bundle over $X$ and $\mu_D(\CL)$ is the first
Chern class of the determinant line bundle \cite{Gottsche:2006bm,
  Gottsche:2019vbi}. This invariant can also be refined to the $\chi_y$
genus. Since (\ref{chihol}) is a measure for the number of sections of
 $\CL$, it is also known as Verlinde
number in analogy with two-dimensional conformal field theory \cite{Losev:1995cr}.
  
Ref. \cite{Gottsche:2006bm} determines generating functions for the
complex projective plane $\mathbb{P}^2$. The results further
strengthen the notion of ``strange duality'' between the instanton
number and the degree of $\CL$ \cite{marian2008tour}. 
For manifolds with $b_2^+>1$, there is much evidence that the generating functions can be expressed as
products of universal functions \cite{Gottsche:2019vbi}, similar to the case of Euler
characteristics discussed above. 

\subsubsection*{Segre numbers}
These numbers are intersection numbers of Chern classes of virtual
bundles over the moduli spaces \cite{Gottsche:2020ass}. Let $\alpha$ be a
  K-theory class of $X$. Then there is a canonical way to obtain a
  K-theory class $\alpha_{\CM_k}$ of $\CM_k$ \cite{Gottsche:2020ass}. We set
  \be
\label{SegreNumber} 
  S(k,\alpha)=\int_{[\CM_k]^{\rm vir}} c(\alpha_{\CM_k}).
\ee
Generating functions for these invariants can also be expressed in
terms of products of universal functions \cite{Gottsche:2020ass}. Moreover, there is an explicit correspondence between
the Verlinde numbers and Segre numbers, which was first studied for
the Hilbert scheme of points \cite{Johnson_2018, marian2021higher}.
\subsubsection*{Elliptic genera of $\CM_k$}
We can moreover consider the elliptic genus of $\CM_k$. For $X=K3$, the result is known since
  the 1990's as the inverse of the Igusa cusp form \cite{Dijkgraaf:1996it, Dijkgraaf:1996xw}. Generating functions for elliptic genera of
  moduli spaces of rank 2 sheaves are proposed in \cite{Gottsche:2018epz}.    
 
\subsubsection*{Family Donaldson invariants}    
 Donaldson invariants for smooth families of four-manifolds were suggested by
 Donaldson in \cite{Donaldson1996} and developed further in
 \cite{li2001}. Here one considers a fibre bundle $\CX\to B$ with
 fibre a four-manifold $X$, and analogues of instanton and monopole equations on
 $\CX$. The family invariants are then obtained through the
 intersection theory on the corresponding moduli spaces, and can be
 viewed as cohomology classes on ${\rm BDiff}(X)$. An interesting aspect is that these numbers are sensitive to
wall-crossing if $b_2^+(X)={\rm dim}(B)+1$.

\subsubsection*{Bauer-Furuta invariants}
These invariants are valued in a cohomotopy group \cite{Bauer2004}, and are more refined
than SW invariants. For example, they can distinguish among direct sums, for which the SW
invariants vanish. See the review \cite{BAUER2006457} for more details.


\section{Topologically twisted Yang-Mills theory}
\label{TTYM}

This section gives a brief introduction to topologically twisted
quantum field theory. Subsection \ref{GenAsp} gives an introduction to the
elementary aspects, and Subsection \ref{TwistPro} reviews the most well-known
topologically twisted Yang-Mills theories and their $Q$-fixed
equations. 

\subsection{General aspects}
\label{GenAsp}
The characteristic feature of a topologically twisted field theory is
that it results from a certain operation, known as topological
twisting, applied to a supersymmetric field theory such that the global
symmetry group of the resulting theory contains a scalar, nilpotent
supercharge $Q$ with $Q^2=0$. Since correlation functions of a $Q$-exact
observable vanish (formally) \cite{Witten:1988ze}, it is natural to
consider the set of $Q$-closed observables, namely those observables $\CO$ such
that $Q\CO=0$. Moreover for a topologically twisted theory, the metric
dependence of the Lagrangian is $Q$-exact, such that correlation functions of $Q$-closed
observables are independent of the metric.

As a result, the interest for four-manifold
topology is in correlation functions of observables in the
$Q$-cohomology $H_Q$,
\be
\left< \CO_1(\vec x_1)\dots \CO_n(\vec x_n) \right>,\qquad \CO_j\in H_Q.
\ee 
Being independent of the metric, these correlation functions are
independent of the positions $\vec x_j$. Yet, the correlation
functions may depend on the smooth structure on $X$, which is required for
the formulation of the theory.

Topological twisting of a Yang-Mills theory with $\mathcal{N}=2$ or $\mathcal{N}=4$
supersymmetry amounts to the specification of a principal
bundle for the global symmetry group, including the $R$-symmetry group
$G_R$ and possibly the flavor group $G_F$ acting on the hypermultiplets. Considering first the case that
$X$ is spin and $G_F$ trivial, the principal bundle is specified by a
homomorphism \cite{Vafa:1994tf, Kapustin:2006pk} 
\be
\label{twisthom}
\phi:{\rm Spin}(4)\to G_R.
\ee
We furthermore set ${\rm Spin}'(4)=(1\times\phi) {\rm Spin}(4)\subset {\rm
  Spin}(4)\times G_R$. Let $S^+$ be the spin
representation of ${\rm
  Spin}(4)\times G_R$ of the supersymmetry generators on $\mathbb{R}^4$. For a topologically twisted theory, we require
that the ${\rm Spin}'(4)$ representation of $S^+$ contains a singlet $Q$,
and that the associated bundle to this representation exists for any
oriented four-manifold $X$.

\subsection{The prototypical twists of $\CN=4$ supersymmetry}
\label{TwistPro}
Having discussed the general aspects of a twisted theory, we continue
by discussing the prototypical examples of $\CN=4$ Yang-Mills. 
This theory is superconformal and uniquely determined by its gauge
group and ultraviolet coupling constant $\tau_{\rm uv}$. Moreover, the
$R$-symmetry group $G_R=SU(4)$ and the flavor group is trivial. There
are three inequivalent choices for the topological twist
\cite{Yamron:1988qc} identified by specifying the action of
$SU(2)_+\times SU(2)_- \simeq {\rm Spin}(4)$ on the
4-dimensional representation ${\bf 4}$ of $SU(4)$:
\begin{itemize}
\item Donaldson-Witten twist: the ${\bf 4}$ of $SU(4)$  transforms as
  $({\bf 1},{\bf 2})\oplus ({\bf 1},{\bf 1}) \oplus ({\bf 1},{\bf 1})$,
\item Vafa-Witten twist: the ${\bf 4}$ of $SU(4)$
  transforms as $({\bf 1},{\bf 2})\oplus ({\bf 1}, {\bf 2})$,
\item Kapustin-Witten twist: ${\bf 4}$ of $SU(4)$
  transforms as $({\bf 2},{\bf 1})\oplus ({\bf 1}, {\bf 2})$.
\end{itemize}
We briefly review these different twists:

\subsubsection*{Donaldson-Witten twist}
The bosonic fields of the Donaldson-Witten twist of the $\CN=4$
Yang-Mills theory are a field strength $F$, a complex scalar $\phi$, and a bosonic complex spinor
$M_{\dot \alpha}$, a section of the anti-chiral spin bundle. 
Thus as stated, this theory is only well-defined if $X$ is a spin
manifold. The twist preserves the subgroup $U(1)_B \subset SU(4)$,
which acts on the spinors $M_{\dot \alpha}$.  The $Q$-fixed equation of this theory is the non-Abelian
version of \eqref{SWeqs} in the adjoint representation \cite{Witten:1994cg, Labastida:1995zj, Labastida:1998sk}.
The non-Abelian monopole equation has multiple fixed loci under the
$U(1)_B$ action:
\begin{itemize} 
\item the instanton component $\CM^{\rm i}_k$. Here $M_{\dot \alpha}=0$, and the
  equation reduces to the anti-self duality equation, $F=-*F$. 
\item the monopole or abelian component $\CM^{\rm a}_k$. The action of $U(1)_B$ is pure gauge for
  this component.
\end{itemize}

\subsubsection*{Vafa-Witten twist} The bosonic fields of this theory are a gauge
  field $F$, a scalar $C$ and a self-dual two-form $B_{\mu\nu}$
  \cite{Vafa:1994tf}. Thus this theory can be formulated on any
  compact, oriented manifold, not necessarily spin. The $Q$-fixed equations are
  \be
  \label{VWeqs} 
  \begin{split}
  &  F_{\mu\nu}^++\frac{1}{2}[C,B_{\mu\nu}^+]+\frac{1}{4}[B_{\mu\rho}^+,B_{\nu\sigma}^+]g^{\rho\sigma}=0,\\
& D_\mu C+D^\nu B_{\mu\nu}^+=0.
  \end{split} 
  \ee
Similarly to non-Abelian monopole equations, the solution space to
this equation also has an instanton and monopole component. These equations famously lead to the Euler characteristic of instanton
moduli spaces \eqref{EulerNumber} \cite{Vafa:1994tf,
  Dijkgraaf:1996tz}. The partition function becomes a generating
function of Euler characteristics, say for $G=SU(2)$,
\be
\label{ZVW} 
Z^{SU(2)}_\bfmu(\tau_{\rm uv}) =\sum_n c(n)\, q_{\rm uv}^n,\qquad
q_{\rm uv}=e^{2\pi i \tau_{\rm uv}}. 
\ee  
Electric-magnetic duality or $S$-duality implies modular
transformation properties of the $Z^{SU(2)}_\bfmu$ under
$SL_2(\mathbb{Z})$ transformations of $\tau_{\rm uv}$.
As a result, the generating functions (\ref{ZVW}) are expected to
transform as a modular form under $S$-duality \cite{Vafa:1994tf}. 
  
\begin{figure}[h!]
 \begin{center}
\begin{tikzpicture}[inner sep=2mm,scale=1.6]
 \node (1a) at (0,8) [circle,draw, ultra thick] {$\left[Z_{\bfmu}^{SU(2)}\right]$}; 
 \node (2a) at (0,6) [circle,draw, ultra thick] {$\left[Z_{\bfmu}^{SO(3)_+}\right]$}; 
 \node (3a) at (2,4) [circle,draw, ultra thick] {$\left[Z_{\bfmu}^{SO(3)_-}\right]$};
 \node (3b) at (-2,4) [circle,draw, ultra thick] {$\left[Z_{\bfmu+\bar w_2/2}^{SO(3)_-}\right]$};
 \node (2b) at (0,2) [circle,draw, ultra thick] {$\left[Z_{\bfmu+\bar w_2/2}^{SO(3)_+}\right]$}; 
 \node (1b) at (0,0) [circle,draw, ultra thick] {$\left[Z_{\bfmu+\bar w_2/2}^{SU(2)}\right]$}; 
\drawloop[->,stretch=1.0]{1a}{-50}{50} node[pos=-2,right]{$T$};
\drawloop[->,stretch=1.0]{1b}{-50}{50} node[pos=-2,right]{$T$};
\draw [<->] (1a) to node[auto] {$S$} (2a);
\draw [->] (2a) to node[above] {$T$} (3a);
\draw [->] (3a) to node[below] {$T$} (2b);
\draw [->] (2b) to node[below] {$T$} (3b);
\draw [->] (3b) to node[above] {$T$} (2a);
\draw [<->] (2b) to node[right] {$S$} (1b);
\draw [<->] (3a) to node[above] {$S$} (3b); 
\end{tikzpicture}
\end{center}  
\caption{Action of the generators $S$ and $T$ of $SL(2,\mathbb{Z})$ on
  equivalence classes $[Z^G_\bfmu]$ of partition functions
  $Z^G_\bfmu:=Z_\bfmu[\CT(G)]$. $\bar w_2$ is an integral lift of
  $w_2(X)$ to $L$.}
\label{STPF}
\end{figure}
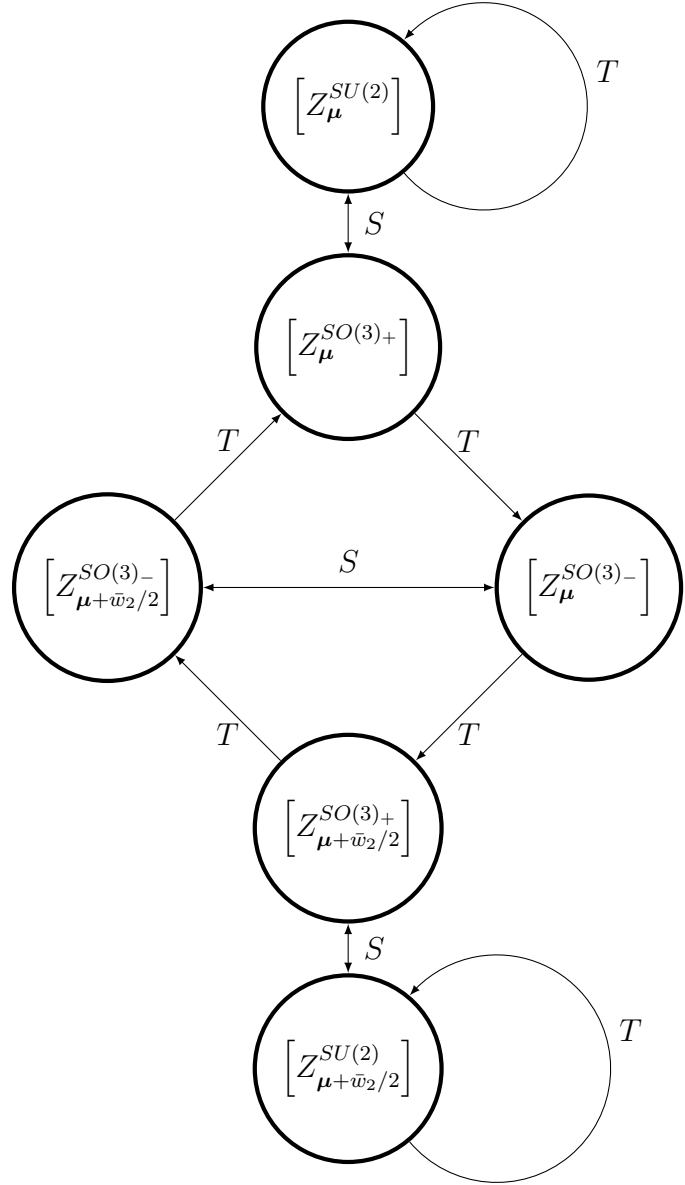

We can gauge the one-form symmetry of this theory in two ways differing
by a discrete theta angle,
\be
\begin{split}
&Z^{SO(3)_+}_\bfmu=\sum_{\bfnu \in (L/2)/L} e^{4\pi i B(\bfmu,\bfnu)} Z^{SU(2)}_\bfnu,\\
&Z^{SO(3)_-}_\bfmu=\sum_{\bfnu \in (L/2)/L} e^{4\pi i
  B(\bfmu,\bfnu)-2\pi i \bfnu^2} Z^{SU(2)}_\bfnu,
\end{split}  
\ee
$SL_2(\mathbb{Z})$ transformations on $\tau_{\rm uv}$ transform the partition functions
$Z^{SU(2)}_\bfmu$, $Z^{SO(3)_+}_\bfmu$ and $Z^{SO(3)_-}_\bfmu$ into
each other as in the Figure \ref{STPF}.  We will elaborate on this in the next section.

A major difficulty to develop this theory on arbitrary four-manifolds is non-compactness of the space of solutions to
\eqref{VWeqs}. See for example \cite{taubes2017behavior}. Fortunately on K\"ahler manifolds and algebraic surfaces, one can work equivariantly with
respect to a $\mathbb{C}^*$ action, which has allowed much progress on
the mathematical side, as reviewed in Section \ref{InvModSpace}. The
partition function (\ref{ZP2}) is then identified with
(\ref{ZVW}) and $q=q_{\rm uv}$. Here a puzzle arises, since (\ref{ZP2}) does not
transform as a modular form, but rather as a (mixed) mock modular
form. It was suggested in \cite{Vafa:1994tf} that the partition function includes
an additional non-holomorphic term such that the result partition
function transforms as a modular form. Indeed, this term was recently
derived as a contribution from the boundary of field space in VW
theory \cite{Dabholkar:2020fde}. This holomorphic anomaly extends to
higher rank and other four-manifolds with $b_2^+=1$
\cite{Minahan:1998vr, Manschot:2017xcr, Dabholkar:2020fde, Alexandrov:2019rth}.

For compact four-manifolds, which can be considered as a gluing of two
four-manifold with boundaries along a three-manifold, the partition
function can be obtained by combining the individual partition functions \cite{Gadde:2013sca}.
For Fano surfaces, Ref. \cite{Beaujard:2020sgs} uses the correspondence between $\CM_k$ and moduli spaces
of quiver representations to determine $\chi(\CM_k)$.

\subsubsection*{Geometric Langlands twist} This twist of $\CN=4$ Yang-Mills was
systemically studied in \cite{Kapustin:2006pk}. The bosonic fields of this theory are the gauge
  field $F$ and a one-form $\phi_\mu dx^\mu$. The $Q$-fixed equations
  depend on an extra parameter $t\in \mathbb{CP}^1$. These equations,
  known as the Kapustin-Witten equations, read
  \be
  \label{KWeqs}
  \begin{split}
&(F-\phi\wedge \phi+t D\phi)^+=0,\\
&(F-\phi\wedge \phi-t^{-1} D\phi)^-=0,\\
&D^*\phi=0.
\end{split}
  \ee
This twist has been very important for connections to the geometric
Langlands program. To this end, the theory is considered on
four-manifolds of the form $X= \Sigma\times C$. The geometric
Langlands program on $C$ can be studied using an effective field
theory on $\Sigma$ \cite{Kapustin:2006pk}. The relation of the
equations (\ref{KWeqs}) to four-manifold geometry is less established. We will not further
discuss this twist here.

\section{$\CN=2$ theories and their twists}
\label{N2theories}
This section discusses theories with $\CN=2$
supersymmetry and their topological twists. With this large family of
physical theories, we can connect to various topological invariants
discussed in Section \ref{InvModSpace}. Many of these connections were
described by Losev {\it et al} \cite{LoNeSha}. The physical
perspective provides new insights for mathematics, while the rigorous
mathematical results also support the phenomena and techniques of
quantum field theory.

The $R$-symmetry group for $\CN=2$ in four dimensions is $SU(2)_R$ which fixes
the twist to be of Donaldson-Witten type, namely the ${\bf 2}$
of $SU(2)_R$ transforms as the $({\bf 1}, {\bf 2})$ representation of
${\rm Spin}(4)$. On the other hand, the hypermultiplets form a quaternionic representation $\mathcal{R}$ of
$G\times G_F\times SU(2)_R$. To ensure that the twisted hypermultiplets can be
formulated on an arbitrary four-manifold, we must specify a homomorphism
\be
\label{TwistHom}
{\rm Spin}^c(4)\to (G\times G_F\times SU(2)_R)/\mathbb{Z}_2,
\ee
compatible with the homomorphism specified above for the vector
multiplets, and such that the representation $\mathcal{R}$ lifts to a proper
bundle over $X$. Different choices of this homomorphism lead to
different topological twists of these theories.
 
The $Q$-fixed equations in the ultraviolet and infrared depend crucially on the details of the
theory and its twisting. The infrared $Q$-fixed equation
can take a different form in different regions of the space of Coulomb
or Higgs vacua. Famously, in the vicinity of a massless monopole on the Coulomb
branch, the $Q$-fixed equation is the abelian monopole equation
\eqref{SWeqs} leading to Seiberg-Witten invariants.

\subsection{Pure $\CN=2$, $SU(2)$ Yang-Mills}
\begin{figure}[ht]\centering 
	\includegraphics[scale=1.2]{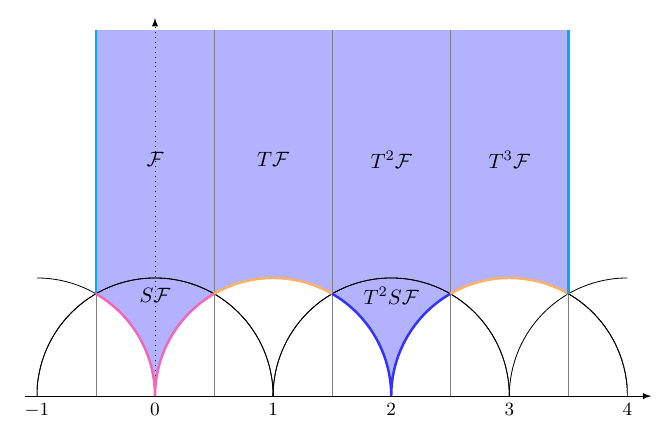}
	\caption{Fundamental domain of $\Gamma^0(4)$.  The two cusps
          on the real line correspond to the strong coupling
          singularities of the gauge theory, while the cusp at
          $\tau=i\infty$ corresponds to weak coupling. The boundaries with the same colour are identified.  }\label{fig:fundgamma0(4)}
\end{figure}

This theory can be considered as the minimal Yang-Mills theory with
$\CN=2$ supersymmetry. It only consists of the vector multiplet,
which includes a gauge potential $A_\mu$, two Weyl
fermions $\psi^I_\alpha, I=1,2$, and a complex scalar $\phi$. The effective coupling
$\tau \in \mathbb{H}$ of the low energy effective theory depends on the order parameter $u$,
\be
u= \frac{1}{16\pi^2} \left<\Tr[\phi^2]\right>_{\mathbb{R}^4},
\ee
where the trace
is in the 2-dimensional representation of $SU(2)$. The order
parameter $u$ can in turn be expressed as function of $\tau$ using the non-perturbative solution of the theory in terms
of the SW curve \cite{Seiberg:1994rs, Seiberg:1994aj}. In terms of
Jacobi theta series $\vartheta_j$, one has
\be
u(\tau)=\Lambda^2 \frac{\vartheta_2(\tau)^4+\vartheta_3(\tau)^4}{2 \vartheta_2(\tau)^2 \vartheta_3(\tau)^2},
\ee
with $\Lambda$ the dynamically generated scale of the theory. The function $u(\tau)$ is a Hauptmodul
for the congruence subgroup $\Gamma^0(4)\subset SL_2(\mathbb{Z})$. The fundamental
domain for $\mathbb{H}/\Gamma^0(4)$ is displayed in Figure \ref{fig:fundgamma0(4)}.
The SW solution was an important step for the development of the relations with integrable systems \cite{Donagi:1995cf, LoNeSha}. See
\cite{Shatashvili:1999} for a concise review.

The Donaldson-Witten twist of this theory transforms the pair of spinor fields
$\psi^I_\alpha$, to a fermionic scalar $\eta$, vector field $\psi_\mu$
and self-dual 2-form $\chi$, while leaving the bosonic fields
unchanged. The twisted fields are the crucial ingredients for the
connection between the Donaldson polynomial invariants \eqref{Dint} and correlation functions
in gauge theory \cite{Witten:1988ze}. The Donaldson differential forms of Section \ref{GTIFM} correspond to
physical observables \cite{Witten:1988ze},
\be
\label{muDObs}
\begin{split}
&  \mu_D(r) \longleftrightarrow 2\,|r|\,u,\qquad r\in H_0(X,\mathbb{Z}),\\
&  \mu_D(\bfx) \longleftrightarrow  I(\bfx)=\frac{1}{4\pi^2}\int_\bfx
  {\rm Tr}\left[\frac{1}{8}\psi\wedge \psi -\frac{1}{\sqrt{2}}\phi F
  \right],\qquad \bfx \in H_2(X,\mathbb{Z}).
\end{split}
\ee
The descent formalism ensures that the observables on the right hand
side are $Q$-closed. The intersection numbers (\ref{Dint}) are
identified with correlation functions of these observables in the
twisted Yang-Mills theory.

The duality properties of the low energy effective theory famously led to a dual
description in terms of Abelian monopole equations and SW invariants
\cite{Witten:1994cg}. Section \ref{Secuplane} sketches the $u$-plane integral
for manifolds with $b_2^+=1$, which is a crucial part the link
between Yang-Mills theory and Donaldson invariants for such four-manifolds.

\subsection{$\CN=2^*$ Yang-Mills}
\label{N=2*YM}
This is the name of the $\CN=2$ theory with the vector multiplet and a
hypermultiplet with mass $m$ in the adjoint representation. The
hypermultiplet consists of two complex bosons, $q,\tilde q$, and two
Weyl fermions $\chi_\alpha,\lambda_\alpha$. This
 theory is the $\CN=2$ preserving mass deformation of the $\CN=4$ theory discussed in Section
 \ref{TwistPro}, and its flavor group $G_F$ is $U(1)_B$. Besides the mass and UV
coupling $\tau_{\rm uv}$, this theory has an effective, infrared coupling $\tau$. Using the SW solution of this theory
\cite{Seiberg:1994aj}, the order parameter $u$ can be expressed as a
bimodular form in $\tau$ and $\tau_{\rm
  uv}$ \cite{Labastida:1998sk, Huang:2011qx}. The corresponding fundamental domain
$\CF^*$ for $\tau$ is the modular curve for the congruence subgroup
$\Gamma(2)$ with the point $\tau_{\rm uv}$ removed.

The twisted fields of the $\CN=2^*$ theory form the same Spin$(4)$
representations as the fields of the Donaldson-Witten twist of $\CN=4$ discussed in Section \ref{TwistPro}.
Specification of the homomorphism (\ref{TwistHom}) involves the
choice of a Spin$^c$ structure $\frak{s}$, with characteristic class
$c(\frak{s})$ \cite{Manschot:2021qqe}. Physically, the choice of Spin$^c$ structure
corresponds to a choice of background flux $\bfk_m=[F_m/4\pi]=c(\frak{s})/2$ for the
weakly gauged $U(1)_B$ symmetry. The $Q$-fixed equations are equal to
the non-Abelian adjoint SW equations. These gives rise to the index
bundle, $W_k\to \CM^{\rm i}_k$ over the instanton moduli space, which is the K-theoretic difference
between the kernel and cokernel bundle, ${\rm Ind}(D_A)={\rm
  Ker}(D_A)-{\rm Coker}(D_A)$. The virtual rank of $W_k$ is
\be
\label{rkW}
{\rm rk}(W_k)=-4k+\frac{3}{8}(c_1(\frak{s})^2-\sigma).
\ee
Combined with (\ref{dimMk}), this gives for the virtual real dimension of the moduli
space of the $Q$-fixed equations, $\CM^Q_k$,  
\be
{\rm vdim}_{\mathbb{R}}(\CM^Q_k)=\frac{3}{4} (c_1(\frak{s})^2-2\chi-3\sigma),
\ee 
which is notably independent of the instanton number $k$. 

After localizing to the instanton and monopole components, the path integral reads \cite{LoNeSha, Manschot:2021qqe}
\be
\label{2StarCorr}
\left< \CO \right>=\sum_k q_{\rm uv}^k \int_{\CM^{\rm i}_k \cup  \CM^{\rm m}_k}
  \frac{1}{{\rm Eul}({\rm Ind}(D_A))}  \wedge \omega_\CO ,
\ee
where ${\rm Eul}({\rm Ind}(D_A))$ is the equivariant Euler class of
the index bundle of the Dirac operator,
\be
{\rm Eul}({\rm Ind}(D_A))=\frac{{\rm Eul}({\rm Ker}(D_A))}{{\rm Eul}({\rm Coker}(D_A))}.
\ee

If $c_1(\frak{s})=2\bfk_m$ corresponds to the canonical Spin$^c(4)$ structure associated to an ACS, as
described in Section \ref{4manBasic}, the
adjoint SW equations are a compact deformation of the Vafa-Witten
equations (\ref{VWeqs}). The index bundle over $W_k$ becomes the tangent bundle
to $\CM_k^{\rm i}$, such that (\ref{2StarCorr}) becomes the generating
function of Euler characteristics of $\CM_k$ for $\CO=1$. The
partition functions of the VW theory and this twist of the $\CN=2^*$ theory
are argued to be identical \cite{Manschot:2021qqe}.

For $\bfk_m=0$, the path integral was studied in
\cite{Labastida:1998sk}. For different values of $\bfk_m$ the path
integral was formulated and evaluated in \cite{Manschot:2021qqe}.
While the path integral is formally holomorphic in $\tau_{\rm uv}$, it
may depend on $\bar \tau_{\rm uv}$ for $b_2^+=1$. This will be
discussed in some more detail in Section \ref{u2Star}.

\subsection{$\CN=2$ SQCD}
$\CN=2$ SQCD refers to the theories with $N_f\leq 3$ hypermultiplets
in the fundamental representation of $SU(2)$. For generic masses
$m_j$, $j=1,\dots, N_f$, the flavor symmetry group is
$U(1)^{N_f}$. The data for the topologically twisting of these
theories includes the choice of background fluxes $\bfk_j=[F_j/4\pi]$
for the flavor bundle \cite{Hyun:1995mb, Hyun:1995hz}. A
consistent formulation requires \cite{Aspman:2022sfj}
\be
\label{SQCDkj}
\bfk_j \in (\bar w_2(X)+\bar w_2(E))/2+L=K_X/2+\bfmu+L,
\ee
with $\bfmu$ as in (\ref{bfmu}). Note that for the fundamental representation the quantization of
$\bfk_j$ depends on the 't Hooft flux $\bfmu$. 
The $Q$-fixed equations
are the non-Abelian monopole equations in the fundamental
representation, possibly with multiple monopole fields.

Similarly to $\CN=2^*$, the Dirac equation gives rise to $N_f$ index
bundles $W^j_k\to \CM^{\rm i}_k$, $j=1,\dots,N_f$ with first Chern class
$c_1(\CL_j)=2\bfk_j$. Their virtual rank is
\be
\label{rkW}
{\rm rk}(W^j_k)=-k+\frac{1}{4}(c_1(\CL_j)^2-\sigma),
\ee
which is an integer for $\bfk_j$ as in (\ref{SQCDkj}). 
Since the hypermultiplets are in the fundamental representation, the
theory does not have a 1-form symmetry associated to the center of $SU(2)$. We can
therefore not gauge this symmetry and sum over 't Hooft fluxes in
$(L/2)/L$. Indeed, these 't Hooft fluxes require different flavor background
fluxes as stipulated in (\ref{SQCDkj}). The virtual dimension of the moduli space of $Q$-fixed equations
 is the sum of (\ref{dimMk}) and (\ref{rkW}),
\be
{\rm vdim}_{\mathbb{C}}(\CM^{Q,N_f}_{k,\CL_j})=(4-N_f)k+\frac{1}{4}\left(
  -3\chi-(3+N_f)\sigma +\sum_{j=1}^{N_f}c_1(\CL_j)^2\right).
\ee

Correlation functions can be expressed similarly to (\ref{2StarCorr})
for $\CN=2^*$, but now with the Dirac operator coupled to
the bundles $W^j_k$. Since the rank (\ref{rkW}) of the index bundle is
negative for sufficiently large $k$, the contribution to $\left< \CO \right>$ from the instanton branch 
becomes \cite{LoNeSha, Manschot:2021qqe}
\be  
\label{SQCDUV}
\sum_k \Lambda_{N_f}^{{\rm 
    vdim}(\CM_k^{Q,N_f})} \int_{\CM_k^i } \omega_{\CO} \wedge \left(\prod_{j=1}^{N_f} m_j^{-{\rm
    rk}(W^j_k)} \sum_l \frac{c_{j,l}}{m_j^l}\right),    
\ee
with $c_\ell$ the Chern classes of ${\rm Coker}(D_A)$. This evaluates
to a polynomial in the masses $m_j$ if ${\rm Coker}(D_A)$ is a proper
bundle, or possibly an
infinite series if it is a sheaf rather than a vector bundle. 
Comparing with (\ref{SegreNumber}), we conclude that topological
correlation functions of $\CN=2$ SQCD evaluate to Segre numbers.

Explicit results for massless SQCD, $N_f=2,3$ were derived in
Ref. \cite{Malmendier:2008db} for $b_2^+=1$, and in Ref. \cite{Kanno:1998qj} for
$b_2^+>1$. For massless, and more generally equal masses, there are SW type
contibutions from multi-monopole equations \cite{bryan1996}. Ref. \cite{Dedushenko:2017tdw} discusses structural results for the mass dependence and
relation to vertex algebras. Ref. \cite{Aspman:2023ate} gives various
explicit results on such generating functions for finite masses, which
confirm the connection to Segre invariants and relations as \eqref{SQCDUV}. With the explicit
expressions, the behavior under special limits, such as decoupling
limits and limits to the AD mass locus, can be analyzed.

\subsection{$\CN=1$ Yang-Mills on $X\times S^1$}
To connect to the K-theoretic invariants in Section \ref{InvModSpace}, we
introduce the five-dimensional theory with $\CN=1$
supersymmetry \cite{Seiberg:1996bd, Intriligator:1997pq}. The bosonic field content is a gauge field $A_m$,
$m=0,\dots 4$ and a real scalar field $\sigma$ with domain the half
line, $\sigma \geq 0$. This theory is invariant under an additional
global symmetry group, $U(1)_I$,
with current 
\be
j=*\left(\frac{1}{8\pi^2}{\rm Tr}\,F\wedge F\right).     
\ee 
The instanton particles are charged under this symmetry. In addition
to the usual kinetic terms, the Lagrangian of this 5-dimensional
theory can include a Chern-Simons (CS)
term
\be
\label{SCS}
S_{\rm CS}=-\frac{i\kappa}{24\pi^2}\int {\rm Tr}\left[ A\wedge F\wedge F\right]+\dots,
\ee
where the $\dots$ represent the terms necessary to make the CS term
supersymmetric. The CS level $\kappa$ is crucial for matching the
results from instanton
counting and with geometric engineering \cite{Tachikawa:2004ur}.

We can moreover weakly gauge the $U(1)_I$ by
introducing a ``frozen'' vector multiplet with bosonic fields $(A^I_m,\sigma^I)$. We can
introduce a mixed CS term \cite[Section 6]{Losev:1995cr, Baulieu:1997nj}, 
\be
S_{\rm mixed\,\,CS}=i\int F^I\wedge {\rm Tr}[AdA+\tfrac{2}{3}A^3]+\dots.
\ee
 
When considered on $\mathbb{R}^4\times S^1$, the theory contains a Kaluza-Klein tower of states
and another global symmetry group $U(1)_{KK}$. The order parameter for
the $SU(2)$ KK theory is the expectation value of the Wilson line,
\be
U=\left< {\rm Tr} P \exp\left[\int_{S^1}(\sigma +iA_5)dx_5\right]\right>.
\ee
Similarly  to the 4d theories, the function $U$ as function of the
effective coupling $\tau$ can be derived from its SW solution studied
in \cite{Nekrasov:1996cz, Gottsche:2006bm, Closset:2021lhd}. The
$U$-plane is a branched double cover of the $u$-plane of the pure
$\CN=2$ theory \cite{Kim:2024}.
      
To connect to four-manifold invariants, we aim to formulate the theory on 
$X\times S^1$ with a partial topological twist on $X$.
Using the reduction of degrees of freedom to $S^1$,
\cite{Losev:1995cr, Nekrasov:1996cz} has argued that the theory determines the
Dirac index of $\CM_k$, which assumes that $\CM_k$ is a spin manifold. We can furthermore
couple to a background flux $n_I\in H^2(X,\mathbb{Z})$ for the $U(1)_I$
symmetry. The partition function then reads \cite{Losev:1995cr, Nekrasov:1996cz}
\be
Z_\bfmu(\mathcal{R})=\sum_k {\rm Ind}(\slashed D_{A}, \CM_k)\,\mathcal{R}^{4k},
\ee
with $A$ the connection of the line bundle with first Chern class
$\mu_D(n_I)$. The relation between the index densities for the
twisted holomorphic Euler character (\ref{chihol}) and ${\rm
  Ind}(\slashed D_{A}, \CM_k)$ then demonstrate that
the 5d theory on $X\times S^1$ with background flux $n_I$ captures
the same information as the twisted holomorphic Euler characteristics
described in Section \ref{InvModSpace}. Ref. \cite{Kim:2024}
formulates and evaluates the topologically twisted theory and
reproduces results from the mathematical literature. 

Similarly, a non-trivial flux for $U(1)_{KK}$ can be considered. The
underlying geometry is then a non-trivial fibration of $S^1$ over
$X$ \cite{Closset:2022vjj}.

\subsection{Twisted theories with gauge groups of rank $>1$}
One can generalize from the gauge groups $SU(2)$ or $SO(3)$ with rank 1 to higher rank gauge groups
\cite{Marino:1998bm}. Donaldson invariants for higher rank bundles are
introduced mathematically by Kronheimer \cite{kronheimer2005}. The correlation functions can be expressed
in terms of products of rank 1 SW invariants times universal functions
\cite{Marino:1998bm}. 
     
\subsection{Superconformal theories} 
The large family of $\CN=2$ theories includes interacting
superconformal theories (SCFT). Some of these theories can be realized as
limits theories discussed above. In particular the Argyres-Douglas (AD)
theories \cite{Argyres:1995jj, Argyres:1995xn} are realized at points in the parameter space where
BPS states with mutually non-local electric-magnetic charges become
massless. For example, the masses in $\CN=2$ SQCD can be chosen on a
locus, such that the Coulomb branch contains the superconformal AD points.
 
While not obvious from the field content and interacting nature of these theories,
Refs \cite{Marino:1998tb, Moore:2017cmm} demonstrated that correlation functions
of these theories are also expressed in terms of SW invariants. Yet,
these theories do give rise to
new results for four-manifold invariants since the correlation functions are potentially divergent
in the limit to the AD mass locus. The physical condition that the limit
is smooth gives rise to sum rules for SW invariants
and the notion of superconformal simple type 
\cite{Marino:1998tb}. Under certain conditions including $b_2^+\geq 3$, it is proven that SW
simple type implies superconformal
simple type \cite{Feehan_2019b}.

\subsection{Family Donaldson invariants}
The family Donaldson invariants are realized physically by coupling
topological twisted Yang-Mills theory to a topological twist and
truncation of conformal supergravity \cite{Cushing:2023rha}.
 This theory results in cohomology
 classes on the space of metrics modulo diffeomorphisms, ${\rm
   Met}(X)/{\rm Diff}(X)$ which is a model for BDiff$(X)$\cite{Cushing:2023rha}.

\subsection{Dictionary} 
\label{MathPhysDict}

We conclude this section with a table summarizing the relations between
the invariants and the physical theories.
\begin{table}[h!]
  \begin{center}
    \renewcommand{\arraystretch}{2.2}
 \begin{tabular}{|l | r| }
\hline 
Twist of: &      Invariant of $\CM_k$: \\
    \hline
$\CN=2$ vector multiplet &     Donaldson invariant \\
 $\CN=2$ vector multiplet with adjoint
                     hypermultiplet &  Euler characteristic  \\
   $\CN=2$ vector multiplet with fundamental
                   hypermultiplets & Segre numbers 
   \\
$\CN=2$ vector multiplet coupled to
                              gravity & Family Donaldson invariant \\
   $\CN=1$ 5d vector multiplet on
                                 $X\times S^1$ &  Holomorphic Euler characteristic  \\
 $\CN=1$ 6d vector multiplet on $X\times T^2$ &   Elliptic genus \\
   \hline
 \end{tabular}
\end{center}
\caption{Table with twists of supersymmetric theories and
  corresponding gauge theoretic invariants.}
\end{table}

\section{$u$-plane integrals for manifolds with $b_2^+=1$}
\label{Secuplane}
For four-manifolds with $b_2^+\leq 1$, there is a non-trivial
contribution to the path integral from the Coulomb branch \cite{Witten:1995gf, Moore:1997pc, LoNeSha}. 
This is the
phase of the theory where the gauge group is broken to its (maximal)
Abelian subgroup. A scaling argument of the metric, $g\to t g$ with
$t\to \infty$, shows that this contribution simplifies dramatically
for the topologically twisted theory. It vanishes for $b_2^+>1$ due to fermionic zero modes
\cite{Moore:1997pc}, while for $b_2^+=1$, the path integral reduces to a
finite dimensional integral over zero modes, known as the $u$-plane integral. \footnote{For $b_2^+=0$, there
  are further contributions at 1-loop, which are hardly explored in the literature.} Its evaluation requires
significant input from physics, in particular the full
non-perturbative solution of the low energy effective theory by Seiberg and Witten
\cite{Seiberg:1994rs, Seiberg:1994aj}.  
While for $b_2^+>1$ there is no contribution from the $u$-plane integral,
the integral still provides a way to determine the
partition functions using the phenomenon of
wall-crossing \cite{Moore:1997pc}. This approach has also manifested a link with another field in
mathematics, namely analytic number theory. 

This section reviews the recent progress on this aspect. After a brief
introduction, we discuss it for a few different theories.

\subsection{Introduction to the $u$-plane integral}
  
The full partition function for $N_f$ generic masses reads for a four-manifold,
\be
Z^J_\bfmu=\Phi^J_\bfmu+\sum_{j=1}^{2+N_f} Z^J_{SW,j,\bfmu},
\ee
where $\Phi_\bfmu$ is the contribution from the $u$-plane, and
$Z^J_{SW,j,\bfmu}$ are SW contributions with delta-function support on
the strong coupling singularities. For manifolds with $b_2^+>1$,
$\Phi_\bfmu$ vanishes. 

For $b_2^+=1$, the $u$-plane integral reads schematically \cite{Moore:1997pc},
\be
\label{uplaneIntegral}
\Phi^J_\bfmu[\CO_1\dots \CO_n]=\int da\wedge d\bar a\, \CO_1\dots
\CO_n\,  A^{\chi} B^\sigma\,  \frac{d\bar \tau}{d\bar a}\, \Psi^J_\bfmu(\tau,\bar \tau).
\ee
where $a$ is a local coordinate on the Coulomb branch, and
$\tau=\tau_{\rm ir}\in \mathbb{H}$ is the low energy effective coupling on the Coulomb branch. $A$ and $B$ are topological couplings, whereas $\Psi^J_\bfmu$ is the sum
over the fluxes of the unbroken $U(1)$ gauge group. It reads
\be
\label{Psi}
\Psi_\bfmu^J(\tau,\bar \tau)=\sum_{\bfk \in \Lambda+\bfmu}
(-1)^{B(\bfk,K)} q^{-\bfk^2_-/2} \bar q^{\bfk^2_+/2},
\ee
with $\bfk_+=B(\bfk,J)\,J$ and $\bfk_-=\bfk-\bfk_+$.
The requirement that the integral is well-defined on the Coulomb branch provides a
non-trivial consisteny check on the various non-perturbative couplings.

Eq. (\ref{uplaneIntegral}) assumes that the observables $\CO_j$ are
independent of the abelian flux $\bfk$. If the observable does depend
on $\bfk$, such as $I(\bfx)$ (\ref{muDObs}), it should be included in
the summand of $\Psi_\bfmu^J$ (\ref{Psi}). Depending on the observables $\CO_j$, the effective field theory may
give rise to further contact terms in the integrand of
\eqref{uplaneIntegral} \cite{Moore:1997pc, LoNeSha}. Through evaluating
$\Phi_\bfmu^J$, Moore and Witten provided a physical derivation of the
Donaldson invariants, and matched explicitly with the results by G\"ottsche using
algebraic-geometric techniques \cite{ellingsrud1995wall,
  Gottsche:1996, Gottsche:1996aoa}.  
 
In recent years, the evaluation of the $u$-plane integral has become
more efficient using a special class of mathematical functions,
known as mock modular forms \cite{Malmendier:2008db,
  Malmendier:2010ss, Korpas:2017qdo, Korpas:2019cwg}. 
To this end, one changes variables in (\ref{uplaneIntegral}) to $\tau$ and $\bar \tau$,
such that the integration domain $\CF$ is a fundamental domain for the effective
coupling $\tau$ under electric-magnetic duality transformations. The
domain for the $SU(2)$ without hypermultiplets is displayed in Figure
\ref{fig:fundgamma0(4)}. More generally, the domain depends
sensitively on the theory and its parameters \cite{Nahm:1996di, Closset:2021lhd, Aspman:2021vhs, Aspman:2021evt}.

Mock modular forms are an interesting generalization of
the concept of a modular form. Their history goes back to Ramanujan
(1887-1920), and they have since appeared at many places in mathematics and
theoretical physics \cite{ZwegersThesis, MR2605321,
  Manschot:2007ha, Manschot:2009ia, Dabholkar:2012nd, Larry}. Rather
than giving their definition, I introduce here a famous example.
Let $H(\tau)$ be defined by
\be
\begin{split}
H(\tau)&=2\frac{\vartheta_2(\tau)^4-\vartheta_4(\tau)^4}{\eta(\tau)^3}-\frac{48}{\vartheta_3(\tau)}\sum_{n=1}^\infty
\frac{q^{\frac{1}{2}n^2-\frac{1}{8}}}{1+q^{n-\frac{1}{2}}}\\
&=2q^{-\frac{1}{8}}(-1+45\,q+231\,q^2+770\,q^3+\dots),
\end{split}
\ee 
where $\vartheta_j$ are Jacobi theta series. This $q$-series is a
(weakly) holomorphic function. Another striking property 
of $H(\tau)$ is that the coefficients are identified as linear
combinations of dimensions of irreducible representations of the finite sporadic Mathieu group $M_{24}$ \cite{Eguchi:2010ej}. 
 We furthermore define the non-holomorphic function
\be
\label{widehatH}
\widehat H(\tau,\bar \tau)=H(\tau)-12 i\int_{-\bar \tau}^{i\infty}\frac{\eta(w)^3}{\sqrt{-i(w+\tau)}}dw.
\ee
The distinguishing property that identifies $H(\tau)$ as a mock
modular form is that $\widehat H(\tau,\bar \tau)$ transforms as a
modular form under $SL_2(\mathbb{Z})$. From Eq. (\ref{widehatH}) follows
\be
\label{barderiva}
\frac{\partial \widehat H(\tau,\bar \tau)}{\partial \bar
 \tau}=-\frac{12i}{\sqrt{2}y} \overline{\eta(\tau)^3}.
\ee 

Another famous example of a
mock modular form is the class number generating function, ie the the
numerator of (\ref{ZP2}) \cite{Zagier:1975}. The partition functions
for higher rank mentioned below (\ref{ZP2}) motivated in part the
development of mock modular forms of higher depth
\cite{Alexandrov:2016enp, Manschot:2017xcr, Bringmann:2018cov, Alexandrov:2020bwg}.

The characteristic property (\ref{barderiva}) of mock modular forms makes it possible to efficiently
evaluate $u$-plane integrals. Using these functions one can determine
suitable functions $\widehat F_\bfmu(\tau,\bar \tau)$ on $\CF$ such that
\be
\frac{\partial \widehat F_\bfmu(\tau,\bar \tau)}{\partial \bar
  \tau}=\Psi^J_\bfmu(\tau,\bar \tau)
\ee
It is then straightforward to evaluate the integral as the
coefficient of $q^0$ plus contributions from other cusps of $\CF$ \cite{Korpas:2017qdo, Korpas:2019cwg}
\be
\Phi_\bfmu[\CO]=\left[ \CO\, \nu(\tau)\, F_\bfmu(\tau)\right]_{q^0}+{\rm other\,\,cusps},
\ee
where $F_\bfmu$ is the holomorphic part of $\widehat F_\bfmu$.
  
As mentioned in Section \ref{TTYM}, correlation functions of $Q$-exact
observables formally vanish. On the other hand, it was encountered in
\cite{Korpas:2019ava} that the $u$-plane integral of such observables
may diverge rather than vanish. Building on \cite{1603.03056},
Ref. \cite{Korpas:2019ava} developed a
regularization for the $u$-plane integral which preserves the BRST
symmetry. Using this approach, intersection numbers can be determined for high
instanton number \cite{Korpas:2019cwg}.

\subsection{Overview of $u$-plane integrals for various theories}

The following gives a brief overview of results based on the $u$-plane
integral for various topologically twisted theories.

\subsubsection*{$\CN=2^*$ Yang-Mills}
\label{u2Star}
The $u$-plane integral of this theory includes additional couplings
compared to (\ref{uplaneIntegral}) due to the coupling to the Spin$^c$
structure described in Section \ref{N=2*YM}. These arise as a background flux $\bfk_m$ for the $U(1)_B$
symmetry, and are determined by weakly gauging this symmetry. This results in a holomorphic and non-holomorphic elliptic variables in the theta series $\Psi_\bfmu^J$ \cite{Manschot:2021qqe}.  The $u$-plane integral is
\be
\label{PhiN=2*}
\Phi^J_\bfmu(\tau_{\rm uv},\bar \tau_{\rm uv},\bfk_m)[\CO]= \int_{\CF^*} d\tau \wedge d\bar \tau \, \CO\,\nu(\tau,\tau_{\rm
  uv}, m)\,\Psi_\bfmu(\tau,\bar \tau,
v \bfk_m, \bar v \bfk_m).
\ee
The integrand is only well defined on the domain $\CF^*$ of the
$\CN=2^*$ theory if $2\bfk_m$ corresponds to a Spin$^c$ structure
\cite{Manschot:2021qqe}. This integral was studied for $\bfk_m=0$ in
Ref. \cite{Labastida:1998sk}. 

For the canonical Spin$^c$ structure, evaluation of $\Phi^J_\bfmu$
reproduces known results of VW theory \cite{Manschot:2021qqe}, as
described in Section \ref{N=2*YM} including the holomorphic anomaly,
and the partition functions for $b_2^+>1$ in terms of universal
functions. Explicit results can be derived for other choices of
$\bfk_m$.

\subsubsection*{$\CN=2$ SQCD}
Similarly to the previous case, the $u$-plane integrand for $\CN=2$
SQCD involves additional couplings due to the background fluxes
$\bfk_j$, $j=1,\dots,N_f$ \cite{Aspman:2022sfj}. 
For vanishing $\bfk_j$, the integral is formulated in
\cite{Moore:1997pc}. Malmendier and
Ono \cite[p. 35/36]{Malmendier:2008db} evaluate the integral 
for $N_f=2,3$ on $\mathbb{P}^2$. Eq. (\ref{SQCDUV})
demonstrates that the physical correlation functions equal
intersection numbers including the top Chern class of the matter
bundle, which are also known as Segre numbers in the mathematics literature.

Ref. \cite{Aspman:2023ate} evaluates the partition function
for non-vanishing masses with and without background fluxes. The
inclusion of the $N_f$ background couplings is very similar to the dependence in
\eqref{PhiN=2*}. In agreement with Eq. (\ref{SQCDUV}), it is
found that these are polynomials in the mass parameters, which agree
with the decoupling limit $m_j\to \infty$ from $N_f$ to $N_f-1$, and
other mass limits.

\subsubsection*{$\CN=1$ on $X\times S^1$}
The $U$-plane integral for this theory is formulated in
\cite{Closset:2022vjj, Kim:2024}. Similarly to the above, the twisting
of this theory contains background fluxes, this times for the $U(1)_I$
symmetry. The coupling to the background flux $n_I$ is
determined by weakly gauging the $U(1)_I$ symmetry. In 
this way, the mathematical results for $\mathbb{P}^2$ by \cite{Gottsche:2006bm} are reproduced. Moreover, using the wall-crossing the results
for $SU(2)$ and $b_2^+>1$ are derived from this physical perspective \cite{Gottsche:2019vbi}.

\subsubsection*{Superconformal theories}
The $u$-plane integral can be formulated for the superconformal
theories. This is useful to deduce the central charges of these
theories from the IR couplings in the integrand  \cite{Shapere:2008zf}. To evaluate the
integrals, one can consider those superconformal theories which can be
realized as a scaling limit of an $\CN=2$ theory with
hypermultiplets. The partition function of the superconformal AD3
theory is explicitly analyzed in this way in Ref. \cite{Moore:2017cmm}.

\section{Concluding remarks}
In the previous sections, we have seen a wide range of subjects in mathematics
and physics. It reflects the rich subject of Donaldson-Witten theory
and four-manifold invariants, and its wide impact. While we included many invariants of moduli
spaces and other topologically twisted theories beyond those of the
original DW theory, we mostly focused on numerical invariants of four-manifolds,
and quantum field theories with known Lagrangians. While not within
the scope of this review, topological twisting of general $\CN=2$ theories, such as in
\cite{Gukov:2017zao}\cite[Section 8]{Manschot:2021qqe}, is clearly an 
interesting and fruitful program.

Moreover, the survey discussed only the most classic example of a
Kaluza-Klein theory obtain from a 5-dimensional theory. The large
set of 5- and 6-dimensional provides opportunities for further
connections with the four-manifold topology, and (potentially) gives rise to more involved four-manifold
invariants such as vertex algebras \cite{Gadde:2013sca, Dedushenko:2017tdw}  and topological modular forms
\cite{Gukov:2018iiq}. 

A related direction worth mentioning is the categorification of
numerical invariants, which has been very successful for
3-manifold topology \cite{donaldson_2002}. Categorification of the
VW invariants leads to the five-dimensional Haydys-Witten equations
\cite{Haydys:2010dv, Witten:2011zz}, which may well yield further insights
for four-manifold geometry.



\end{document}